\documentclass[12pt,a4paper]{article}
\usepackage[utf8]{inputenc}
\usepackage[T1]{fontenc}
\usepackage[english]{babel}
\usepackage{hyperref}
\hypersetup{hidelinks}
\usepackage{relsize}
\usepackage{cancel}
\usepackage{amsmath, amssymb, amsfonts, mathrsfs, bm}
\usepackage{tikz}
\usepackage{braket}
\newcommand{\dd}{\mathrm{d}}

\newcommand{\erfc}{\operatorname{erfc}}
\newcommand{\sgn}{\operatorname{sgn}}
\usepackage{graphicx}
\title{Dynamical Casimir effect in the worldline formulation}
\author{C.~D.~Fosco \\
and \\
B.~C.~Guntsche\\
{\normalsize\it Centro At\'omico Bariloche and Instituto Balseiro}\\
{\normalsize\it Comisi\'on Nacional de Energ\'\i a At\'omica}\\
{\normalsize\it R8402AGP Bariloche, Argentina.}}
\begin{document}
\date{}
\maketitle
\begin{abstract}
We evaluate the effective action for the Dynamical Casimir Effect (DCE) for a real scalar field in $d+1$ dimensions within the worldline formulation of quantum field theory. The scalar field is coupled to a spacetime-dependent mass term, which here plays the role of the moving medium and imposes imperfect boundary conditions on time-dependent surfaces. Expanding in powers of the departure of the geometry from a planar configuration, the worldline path integral factorizes into simpler, lower-dimensional ones. In the limit of a strong coupling to the surface, we recover the Dirichlet result and derive the systematic corrections in inverse powers of the coupling, calculating the imaginary part of the effective action up to arbitrary order of said powers. Finally, we also apply the method to a two-surface configuration.
\end{abstract}
\section{Introduction}
The Dynamical Casimir Effect (DCE) refers to the production of particles from
the vacuum, due to a time dependence of the boundary conditions or even of
background fields~\cite{DCE_Review,Moore,Davies_Fulling}.
While traditionally approached via canonical or functional integral
quantization, the worldline formalism offers a powerful alternative. 
This has indeed been the case for the static Casimir effect, where in the approach popularized by Gies et al.~\cite{Gies_WL} the interaction with the medium is introduced through a background potential. This, in turn, allows for the 
evaluation of the one-loop effective action in the context of the 
worldline~\cite{Worldline_Review}.
Since the background potentials are assumed there to be time independent, the corresponding effective action is proportional to the Casimir energy. We show here how the worldline approach may also be useful for the analysis of 
the DCE. In particular, it allows us to evaluate the effective action as a 
measure of the amplitude for the creation of field quanta out of the vacuum.  

This paper is organized as follows. In Sect.~\ref{sec:themodel} we define the system under consideration and introduce the notation and conventions. In Sect.~\ref{sec:eff} we apply the worldline formalism to the effective action, develop its perturbative expansion, and analyze its dependence on $\lambda$, the coupling to the moving medium, calculating explicitly the imaginary part of the effective action to arbitrary powers of $1/\lambda$. We also identify the general structure of the higher-order terms.

Sect.~\ref{sec:twosurfaces} extends the formalism to a two-surface 
configuration. Finally, Sect.~\ref{sec:conclusions} presents our conclusions.

\section{The model}\label{sec:themodel}
We consider a real scalar field $\varphi(x)$ in $D = d+1$ dimensions,
coupled to dynamical media, in such a way that its Euclidean action ${\mathcal 
S}(\varphi, V)$ may be written, after an integration by parts,
as follows:
\begin{equation}
{\mathcal S}(\varphi, V) \;=\; \frac{1}{2} \, \int \dd^Dx \, \varphi(x) 
\big(-\partial^2 +  V \big)  \, \varphi(x)  \;,
\end{equation}
where $x$ has been used to denote the $d+1$ spacetime coordinates. 
We use the notation $x \equiv (x_\parallel, x_d)$, $x_\parallel \equiv 
(x_0, x_1, \dots, x_{d-1})$ being the spacetime coordinates which
parametrize the world-volume of the dynamical medium. Space–time indices are 
denoted by letters from the middle of the Greek 
alphabet ($\mu,\nu,\ldots$), taking values $0,1,\ldots,d$, with $0$ 
reserved for  the imaginary time (natural units
are used: $\hbar\equiv 1$ and $c\equiv 1$). The metric reduces to 
the $(d+1)\times(d+1)$ identity matrix, so no distinction is implied by 
placing an index upstairs or downstairs. Unless stated otherwise, we also adopt
the Einstein summation convention over repeated indices in monomial expressions.

The role of the potential is straightforward: it suppresses the scalar field wherever it becomes large and positive, and in the strong-coupling limit it enforces Dirichlet boundary conditions. This occurs even though the form of the action places the system within the regime of linear response. Moreover, when more than one object is coupled to the vacuum field $\varphi$, the full potential $V$ is obtained by adding the corresponding contributions from each object, as dictated by the assumed linear response of the media. The additivity of potentials does not, however, imply any linear superposition principle for the physical effects: energies, probabilities, and related observables are not obtained by summing the contributions of the individual parts of the media, but are instead nontrivial functionals of $V$.
	
The potential $V(x)$, for the case of a single surface $\Sigma$, is 
assumed to have the form
\begin{equation}\label{eq:defv}
    V(x) \;=\;  v\big[ F(x)\big] \;\;,\;\;\;
   F(x) \, \equiv \, x_d - \psi(x_\parallel)
\end{equation}
where $\Sigma$ is defined by $x_d = \psi(x_\parallel)$, and $\psi$ is 
treated as a small perturbation around a flat configuration. On the 
other hand, $v$ is a real non-negative function of a single variable, highly 
concentrated around $0$. In fact, the first example we consider 
corresponds to a $\delta$-function (zero-width surface). We find it convenient, 
at this point, to mention that the choice of $F$ is of course not unique. 
Let $F: \mathbb{R}^D \to \mathbb{R}$ be a smooth function with $\nabla F 
\neq 0$ on the level set $\Sigma = F^{-1}(0)$, which defines $\Sigma$. 
By the coarea formula, 
\begin{equation}
	\delta(F(x)) = \frac{1}{|\nabla F(x)|} \, \delta_\Sigma(x),
\end{equation}
where $\delta_\Sigma$ denotes the delta distribution intrinsic to $\Sigma$. It 
follows that the combination $|\nabla F(x)|\,\delta(F(x))$ defines a measure on 
$\Sigma$ that is manifestly independent of the choice of defining function $F$.
For the Monge parametrization $F(x) = x_d - \psi(x_\parallel)$, one has 
$|\nabla F| = \sqrt{1 + (\nabla_\parallel \psi)^2} = 1 + O(\psi^2)$, 
so the discrepancy arising from the $|\nabla F|$ factor enters only at 
second order in derivatives of $\psi$.  Moreover, as explained 
in~\cite{FG_Dirichlet}, this factor contributes only to the real 
(non-dissipative) part of the effective action and is therefore 
irrelevant for the evaluation of pair-creation effects. 

\section{Effective action}\label{sec:eff}
The one-loop effective action $\Gamma(V)$ is defined in terms of the 
vacuum-to-vacuum Euclidean transition amplitude, ${\mathcal Z}$, via
\begin{equation}
e^{-\Gamma(V)}\,\equiv\, \frac{{\mathcal Z}(V)}{{\mathcal Z}(0)}\;,
\;\;\; {\mathcal Z}(V) \,\equiv \,\int {\mathcal D}\varphi \, 
e^{-{\mathcal S}(\varphi, V)} \;, 
\end{equation}
where the ${\mathcal Z}(0)$ factor has been introduced in order to subtract the
usual zero-point contribution of free space (namely, in the absence of any
object). Note, however, that we perform further subtractions to the effective
action, corresponding to terms that do not contribute to pair creation, which is the main focus of this work. 

Since the functional integrals are Gaussian, we can safely apply the worldline
approach~\cite{Gies_WL,Worldline_Review}, whereby we may write
\begin{equation}\label{eq:defwl}
 \Gamma(V) = -\frac{1}{2 (4 \pi)^{\frac{d+1}{2}}} \int_{0^+}^\infty 
\frac{\dd T}{T^{\frac{d+3}{2}}} \int \dd^{d+1}x \left[ \langle e^{-\int_0^T \dd\tau 
 V(x(\tau)) } \rangle_x  -1  \right]
\end{equation}
where
\begin{equation}
V(x(\tau)) \;\equiv\; v\!\big(x_d(\tau) - \psi(x_\parallel(\tau))\big)\;,
\end{equation}
and $\langle \dots \rangle_x$ denotes a functional average over closed loops
$x_\mu(\tau)$ starting and ending at the same point, $x_\mu(0)=x_\mu(T) \equiv 
x$:
\begin{equation}
\langle \ldots \rangle_x \;\equiv\; 
\frac{\int^{x(T)=x}_{x(0)=x} {\mathcal D}x \ldots e^{-\frac{1}{4}\int_0^T \dd\tau 
\dot{x}_\mu(\tau)\dot{x}_\mu(\tau)}}{\int^{x(T)=x}_{x(0)=x} {\mathcal D}x 
e^{-\frac{1}{4}\int_0^T \dd\tau \dot{x}_\mu(\tau)\dot{x}_\mu(\tau)}}
\end{equation}

Expression~\eqref{eq:defwl} for the effective action, when rotated back to 
Minkowski spacetime in its dependence on $\psi$, will allow us to extract the 
dissipative effects due to pair creation from the evaluation of its imaginary 
part. 

\subsection{Perturbative Expansion}\label{sec:perturbative}
Let us expand the potential in powers of $\psi$ 
\begin{equation}
V(x) \,\equiv\, v(x_d - \psi(x_\parallel)) = v(x_d) 
+ \sum_{n=1}^\infty \frac{(-1)^n}{n!} v^{(n)}(x_d) \, 
\Big(\psi(x_\parallel) \Big)^n\;,
\end{equation}
and introduce it into the worldline, to get the respective expansion for the 
effective action: 
\begin{equation}
\Gamma(V) = \Gamma_0 + \Gamma_1(\psi) + \Gamma_2(\psi) + \dots \;.
\end{equation}
A key point is that, in the perturbative expansion above, all required worldline path integrals factorize into parallel and perpendicular averages. Let us display this explicitly for the first few terms.

At zeroth order one obtains
\begin{equation}\label{eq:gamma_0}
	\Gamma_0 \;=\; -\frac{1}{2 (4 \pi)^{\frac{d+1}{2}}} \int_{0^+}^\infty 
	\frac{\dd T}{T^{\frac{d+3}{2}}} \int \dd^{d+1}x \left[ \langle e^{-\int_0^T 
	\dd\tau\, v (x_d(\tau)) } \rangle_{x_d}  -1  \right] \;,
\end{equation}
with
\begin{equation}
\langle \ldots \rangle_{x_d} \;\equiv\; 
		\frac{\int^{x_d(T)=x_d}_{x_d(0)=x_d} {\mathcal D}x_d \ldots 
		e^{-\frac{1}{4}\int_0^T \dd\tau 
		\big(\dot{x}_d(\tau)\big)^2}}{\int^{x_d(T)=x_d}_{x_d(0)=x_d}
		 {\mathcal D}x_d \, e^{-\frac{1}{4}\int_0^T \dd\tau 
		\big(\dot{x}_d(\tau) \big)^2}} \;.
\end{equation}
It is evident from~\eqref{eq:gamma_0} that this contribution to the effective action is divergent, since it is proportional to the spacetime volume swept by the surface over an infinite time interval. We therefore introduce a spacetime box with total evolution time $\mathcal{T}$ and a $(d-1)$-dimensional ``area'' $L^{d-1}$ for the flat surface. It is then natural to evaluate 
${\mathcal E}_0$, the vacuum energy per unit area, in the limit when the size 
of the box tends to infinity: 
\begin{equation}\label{eq:cale_0}
	{\mathcal E}_0 \equiv \Big[\frac{\Gamma_0}{\mathcal{T}\, L^{d-1}} 
	\Big]_{\mathcal{T},\, L \to \infty} 
	\;=\; -\frac{1}{2 (4 \pi)^{\frac{d+1}{2}}} \int_{0^+}^\infty 
	\frac{\dd T}{T^{\frac{d+3}{2}}} \int \dd x_d \left[ \langle e^{-\int_0^T 
		\dd\tau\, v (x_d(\tau)) } \rangle_{x_d}  -1  \right] \;,
\end{equation}
which, depending on $d$ and $v$, is usually divergent in the UV. On the other 
hand, note that it does not contribute to the imaginary part of the effective 
action.

We now consider the first-order term, which after standard manipulations may be written as
\begin{align}\label{eq:gamma_1}
\Gamma_1(\psi) \;= & \frac{1}{2 (4 \pi)^{\frac{d+1}{2}}} \int_{0^+}^\infty 
\frac{\dd T}{T^{\frac{d+3}{2}}} \int_0^T \dd\tau \Big[ 
\int \dd^dx_\parallel \langle \psi(x_\parallel(\tau)) 
\rangle_{x_\parallel} \nonumber\\
& \times \int \dd x_d \langle v'(x_d(\tau) ) 
e^{-\int_0^T \dd\tau' v (x_d(\tau'))}\rangle_{x_d}  \Big] \;.
\end{align}
To tackle this term, it is convenient to deal first with the factor 
involving a functional average of $\psi$. In terms of the Fourier transform of 
$\psi$, \mbox{$\widetilde{\psi}(k_\parallel) \equiv \int 
\dd^dx_\parallel\, e^{-i k_\parallel \cdot x_\parallel} 
\psi(x_\parallel)$}, and
taking into account the fact that the worldline correlation function is:
\begin{equation}
\langle x_\mu(\tau_1) x_\nu(\tau_2) \rangle \;=\; \delta_{\mu\nu} 
\,\Delta(\tau_1 - \tau_2) \;,\;\;\Delta(\tau_1 - \tau_2) \,=\,\frac{(\tau_1 - 
\tau_2)^2}{T} - |\tau_1 - \tau_2|  \;, 
\end{equation}
we find:
\begin{equation}
\int \dd^dx_\parallel \langle \psi(x_\parallel(\tau)) 
\rangle_{x_\parallel} = \widetilde{\psi}(0) =
\int \dd^dx_\parallel \psi(x_\parallel)\;.
\end{equation}
 This term may therefore be set to zero by a shift of origin, chosen so that the mean value of $\psi$ vanishes. We assume that such a choice has been made, and therefore $\Gamma_1(\psi)=0$.

For the second-order term $\Gamma_2(\psi)$ one finds a natural decomposition into two contributions, $\Gamma_2=\Gamma_{2,1}+\Gamma_{2,2}$, namely
\begin{align} \label{eq:gamma22_one_surface}
\Gamma_{2,1}(\psi)
&= \frac{1}{4 (4\pi)^{\frac{d+1}{2}}}
\int_{0^+}^\infty \frac{\dd T}{T^{\frac{d+3}{2}}}
\int_0^T \dd\tau
\nonumber\\
&\quad\times \int \dd x_d \, \langle v''(x_d(\tau)) \rangle_{x_d}
\int \dd^dx_\parallel \,
\langle \big(\psi(x_\parallel(\tau))\big)^2 \rangle_{x_\parallel}
\nonumber\\[2pt]
\Gamma_{2,2}(\psi)
&= - \frac{1}{2(4\pi)^{\frac{d+1}{2}}}
\int_{0^+}^\infty \frac{\dd T}{T^{\frac{d+3}{2}}}
\int_0^T\dd\tau_1 \int_0^{\tau_1}\dd\tau_2
\nonumber\\
&\quad\times
\int \dd^dx_\parallel \,\langle \psi(x_\parallel(\tau_1))
\psi(x_\parallel(\tau_2)) \rangle_{x_\parallel}
\nonumber\\
&\qquad\times
\int \dd x_d \,\langle v'(x_d(\tau_1)) v'(x_d(\tau_2))
 e^{-\int_0^T \dd\tau\,v(x_d(\tau))}\rangle_{x_d}\;.
\end{align}

In both terms we need the average of two $\psi$ insertions, one with coincident proper-time arguments. Using Fourier transforms once again, we obtain
\begin{equation}
\int \dd^dx_\parallel \langle \psi(x_\parallel(\tau_1))  
\psi(x_\parallel(\tau_2)) \rangle_{x_\parallel}\,=\,
\int \frac{\dd^dk_\parallel}{(2\pi)^d} \, 
\big|\widetilde{\psi}(k_\parallel)\big|^2 e^{\Delta(\tau_1-\tau_2) 
k_\parallel^2} \;.
\end{equation}
It then follows that $\Gamma_{2,1}$, independently of the detailed form of $v$, yields a term that cannot generate an imaginary part in the effective action; it may be interpreted as a mass term for $\psi$:
\begin{equation}
\Gamma_{2,1}(\psi) \,=\,\frac{1}{2}m^2\int \dd^dx_\parallel  
\big(\psi(x_\parallel)\big)^2 \;,
\end{equation}
with
\begin{equation}
m^2\,=\,\frac{1}{2 (4\pi)^{\frac{d+1}{2}}} 
	\int_{0^+}^\infty \frac{\dd T}{T^{\frac{d+3}{2}}} \int_0^T \dd\tau 
	\int \dd x_d \langle v''(x_d(\tau)) \rangle_{x_d} \;.
\end{equation}

For the evaluation of $\Gamma_{2,2}(\psi)$, we need to be more explicit about 
the form of $v$. We first assume that $v(x_d) = \lambda \delta(x_d)$.
Inserting the explicit form of the worldline average into~\eqref{eq:gamma22_one_surface} and using the normalization of the free path integral, the relevant object may be written in terms of the propagator of a one-dimensional system:
\begin{align}\label{eq:defi}
I(\tau_1, \tau_2; T) \, \equiv \,\int \dd x_d \langle v'(x_d(\tau_1)) 
v'(x_d(\tau_2))  e^{-\int_0^T 
\dd\tau\, v(x_d(\tau))}\rangle_{x_d}
\;=\;(4 \pi T)^{\frac{1}{2}} \, 	\nonumber\\
\times \int_{x_d(T)=x_d(0)} \mathcal{D}x_d
\,v'(x_d(\tau_1)) v'(x_d(\tau_2)) 
e^{-\int_0^T \big(\frac{1}{4}{\dot{x}_d}^2(\tau) + 
\lambda \delta(x_d(\tau)) \big)\,\dd\tau } \;,
\end{align}
so that $\Gamma_{2,2}$ becomes 
\begin{equation}
\Gamma_{2,2}(\psi)	\,=\,\frac{1}{2} \int \frac{\dd^dk_\parallel}{(2\pi)^d} 
\, \gamma(k_\parallel) 
\big|\widetilde{\psi}(k_\parallel)\big|^2 
\end{equation}
with
\begin{equation} \label{eq:gamma_def}
\gamma(k_\parallel) \,=\, - \frac{1}{(4\pi)^{\frac{d+1}{2}}}  
\int_{0^+}^\infty \frac{\dd T}{T^{\frac{d+3}{2}}} \int_0^T\dd\tau_1 
\int_0^{\tau_1}\dd\tau_2 I(\tau_1, \tau_2; T)\,  
e^{\Delta(\tau_1-\tau_2)k_\parallel^2}\;.
\end{equation}

The value of $\gamma$ at zero momentum corresponds to a redefinition of $m^2$. We shall use this fact to control the UV behavior of the integrals defining $\gamma$.

The path integral~\eqref{eq:defi} may be written in terms of $K$, the 
propagator in the presence of a $\delta$ potential~\cite{GroscheSteiner}, as follows:
\begin{align}\label{eq:Idef}
	I(\tau_1,\tau_2;T) \;=\; (4 \pi T)^{\frac{1}{2}} 
	\lambda^2 \,
	\int_{-\infty}^{+\infty}\!\dd x_d &
	\int_{-\infty}^{+\infty}\!\dd y_d 
	\int_{-\infty}^{+\infty}\!\dd z_d\;\big[	
	K_a(x_d,y_d)\,\delta'(y_d)\nonumber\\
	\times & \, K_b(y_d,z_d)\,\delta'(z_d)\;K_c(z_d,x_d)
\big]
\end{align}
where
\begin{equation}
	K_a \equiv K(T_a;\,\cdot\,,\cdot\,)\,,\qquad
	K_b \equiv K(T_b;\,\cdot\,,\cdot\,)\,,\qquad
	K_c \equiv K(T_c;\,\cdot\,,\cdot\,)\,,
\end{equation}
with proper-time intervals
\begin{equation}\label{eq:times}
	T_a = T - \tau_1\,,\qquad
	T_b = \tau_1 - \tau_2\,,\qquad
	T_c = \tau_2\,,\qquad
	T_a + T_c = T - T_b\,
\end{equation}
and
\begin{align}\label{eq:kernel}
K(t;\,x,y)
&= \frac{1}{2\sqrt{\pi t}}\,e^{-(x-y)^2/(4t)}
\nonumber\\
&\quad
-\frac{\lambda}{4}\,\exp\!\Bigl[\tfrac{\lambda}{2}\bigl(|x|+|y|\bigr)
+ \tfrac{\lambda^2}{4}\,t\Bigr]\;
\erfc\!\biggl[\frac{|x|+|y|+\lambda\,t}{2\sqrt{t}}\biggr],
\end{align}
where $t = \tau'' - \tau'$ denotes the elapsed proper time.
	
Using $\int\!f(x)\,\delta'(x)\,\dd x = -f'(0)$ twice in succession,
the three spatial integrals reduce to four terms involving the
middle kernel $K_b$ and its derivatives evaluated at the origin:
\begin{align} \label{eq:I_reduced}
I
&= (4 \pi T)^{\frac{1}{2}} \lambda^2\!\int\!\dd x_d\Bigl[
\partial_2 K_a(x_d,0)\,\partial_2 K_b(0,0)\,K_c(0,x_d)
\nonumber\\
&\qquad\qquad
+ K_a(x_d,0)\,\partial_1\partial_2 K_b(0,0)\,K_c(0,x_d)
\nonumber\\
&\qquad\qquad
+ \partial_2 K_a(x_d,0)\,K_b(0,0)\,\partial_1 K_c(0,x_d)
\nonumber\\
&\qquad\qquad
+ K_a(x_d,0)\,\partial_1 K_b(0,0)\,\partial_1 K_c(0,x_d)
\Bigr].
\end{align}
where $\partial_1,\partial_2$ denote derivatives with respect to the first and 
second spatial arguments.

Evaluating the derivatives of $K$ at the origin using~\eqref{eq:kernel}, 
performing the $x_d$ integral via the semigroup property of the heat kernel, 
and combining the four terms, we find:
\begin{equation}\label{eq:Ifinal}
I(\tau_1,\tau_2;T) \;=\;\frac{\lambda^2 \sqrt{T}}{2} \left[
\frac{K(T-T_b;\,0,0)}{T_b^{3/2}}
\,+\, \frac{K(T_b;\,0,0)}{(T-T_b)^{3/2}}
\right]
\end{equation}
where $T_b = \tau_1 - \tau_2$.  The expression is manifestly symmetric under
$T_b \leftrightarrow T - T_b$.

Consider now the large-$\lambda$ expansion of $I$. Introducing 
$S \equiv T - T_b$:
\begin{equation}\label{eq:Iasym}
I(\tau_1,\tau_2;T) \;=\;\frac{\sqrt{T}}{\sqrt{4\pi}\,(T_b\,S)^{3/2}}	
\sum_{n=0}^{\infty}\frac{(-1)^n\,(2n+1)!!\,2^n}{\lambda^{2n}}
\left[\frac{1}{S^n}+\frac{1}{T_b^n}\right] \;
\end{equation}
of which the Dirichlet limit ($\lambda\to\infty$) is:
\begin{equation}
I_\infty(\tau_1,\tau_2;T)  
= \frac{\sqrt{T}}{\sqrt{\pi}\,\bigl[T_b(T-T_b)\bigr]^{3/2}} \;.
\end{equation}

\begin{align}
\gamma_\infty(k_\parallel) &=\, - \frac{1}{2\pi(4\pi)^{\frac{d}{2}}}  
\int_{0^+}^\infty \frac{\dd T}{T^{\frac{d+2}{2}}} \int_0^T\dd\tau_1 
\int_0^{\tau_1}\dd\tau_2	\frac{e^{[\frac{(\tau_1 - \tau_2)^2}{T} - |\tau_1 - 
\tau_2|] k_\parallel^2}}{[(\tau_1-\tau_2) (T-\tau_1 + \tau_2)]^{3/2}} 
\nonumber\\
 &=\, - \frac{1}{2\pi (4\pi)^{\frac{d}{2}}}  
\int_{0^+}^\infty \frac{\dd T}{T^{\frac{d+2}{2}}} \int_0^T \dd\tau
	\frac{e^{[\frac{\tau^2}{T} - |\tau|] k_\parallel^2}}{\tau^{3/2} 
	(T-\tau)^{1/2}} \;.
\end{align}
Subtracting its value at $k_\parallel = 0$:
\begin{equation}
	\gamma_D(k_\parallel) 
	\equiv \gamma_\infty(k_\parallel) - \gamma_\infty(0) 
	\,=\, - \frac{1}{2\pi (4\pi)^{\frac{d}{2}}}  
	\int_{0^+}^\infty \frac{\dd T}{T^{\frac{d+2}{2}}} \int_0^T \dd\tau
	\frac{e^{[\frac{\tau^2}{T} - |\tau|] k_\parallel^2} - 1}{\tau^{3/2} 
	(T-\tau)^{1/2}} \;.
\end{equation}
Under a change of variables:
\begin{equation}\label{eq:gammaD_integral}
	\gamma_D(k_\parallel)  
	\,=\, - \frac{1}{2\pi (4\pi)^{\frac{d}{2}}}  |k_\parallel|^{d+2} \,
	\int_{0^+}^\infty \frac{\dd T}{T^{\frac{d+4}{2}}} \int_0^1 \dd\tau
	\frac{e^{T (\tau^2  - \tau)} - 1}{\tau^{3/2} 
		(1-\tau)^{1/2}} \;.
\end{equation}
The double integral can be evaluated in closed form.
Setting $v = \tau(1-\tau)$, the proper-time integral is performed
by dimensional regularization:
\begin{equation}
\int_0^\infty \frac{\dd T}{T^{(d+4)/2}}\;
\big(e^{-v\,T} - 1\big)
\;=\; v^{(d+2)/2}\;\Gamma\!\Big(\!-\tfrac{d+2}{2}\Big)\;,
\end{equation}
where the subtraction is automatically handled by the analytic 
continuation in $d$. The remaining $\tau$-integral yields a Beta function:
\begin{equation}
\int_0^1 \dd\tau\;
\frac{\big[\tau(1-\tau)\big]^{(d+2)/2}}
     {\tau^{3/2}\,(1-\tau)^{1/2}}
\;=\;
\int_0^1 \dd\tau\;\tau^{(d-1)/2}\,(1-\tau)^{(d+1)/2}
\;=\;
\frac{\Gamma\!\big(\tfrac{d+1}{2}\big)\,
      \Gamma\!\big(\tfrac{d+3}{2}\big)}
     {\Gamma(d+2)}\;.
\end{equation}
Combining, one obtains the closed-form expression
\begin{equation}\label{eq:gammaD_closed}
\gamma_D(k_\parallel)
\;=\; -\frac{|k_\parallel|^{d+2}}{2\pi\,(4\pi)^{d/2}}\;
\frac{\Gamma\!\big(\!-\tfrac{d+2}{2}\big)\;
      \Gamma\!\big(\tfrac{d+1}{2}\big)\;
      \Gamma\!\big(\tfrac{d+3}{2}\big)}
     {\Gamma(d+2)}\;,
\end{equation}
which agrees with the result for the Euclidean form factors obtained in 
Refs.~\cite{FG_Dirichlet,FG_Neumann}, where 
the Dirichlet limit was treated as a starting point.
For odd $d = 2q+1$, all Gamma functions are finite and the coefficient 
simplifies to $(-1)^{q+1}(q!)^2/[(2\pi)^{q+1}(2q+1)!(2q+3)!!]$
[cf.\ eq.~\eqref{eq:gamman} with $n=0$].
The first few explicit values are:
\begin{equation}\label{eq:gammaD_explicit}
\gamma_D \;=\;
\begin{cases}
\displaystyle -\frac{1}{6\pi}\;|k_\parallel|^3 & (d=1)\,,\\[8pt]
\displaystyle +\frac{1}{360\pi^2}\;|k_\parallel|^5 & (d=3)\,,\\[8pt]
\displaystyle -\frac{1}{25200\pi^3}\;|k_\parallel|^7 & (d=5)\,.
\end{cases}
\end{equation}

\subsection{Large-\texorpdfstring{$\lambda$}{lambda} corrections to \texorpdfstring{$\gamma$}{gamma}}
Let us now derive the corrections to $\gamma_D$ that arise from keeping 
subleading terms in the $1/\lambda$ expansion of $I$.
Substituting the asymptotic expansion (\ref{eq:Iasym}) into the expression for 
$\gamma$, and performing the center-of-mass integration over the double 
proper-time integral (which produces a factor of $T - T_b$), we find that the 
$n$-th order contribution is:
\begin{align}
\gamma_n(k_\parallel) \,=\, & - 
\frac{(-1)^n\,(2n+1)!!\,2^{n-1}}{\lambda^{2n}\,2\pi\,(4\pi)^{\frac{d}{2}}}
\int_{0^+}^{\infty}\frac{\dd T}{T^{\frac{d+2}{2}}}
\int_0^T \dd T_b\;e^{\Delta(T_b)\, k_\parallel^2}\nonumber\\
& \times
\left[
\frac{1}{T_b^{3/2}\,(T-T_b)^{n+\frac{1}{2}}}
\,+\,
\frac{1}{T_b^{n+\frac{3}{2}}\,(T-T_b)^{1/2}}
\right]\;.
\end{align}
The subtracted form $\gamma_n^{\rm sub}\equiv \gamma_n(k_\parallel) - 
\gamma_n(0)$, after rescaling $T_b = u T$ and $T\to T/k_\parallel^2$, 
reads:
\begin{align}\label{eq:gamman}
\gamma_n^{\rm sub}(k_\parallel) \,=\,  
\frac{(-1)^{n+1}\,(2n+1)!!\,2^{n-1}}{\lambda^{2n}\,2\pi\,(4\pi)^{\frac{d}{2}}}
\, |k_\parallel|^{d+2n+2}\;
{\mathcal C}_n(d)\;,
\end{align}
where
\begin{equation}
{\mathcal C}_n(d) \,=\, \int_{0}^{\infty}\!\frac{\dd T}{T^{\frac{d+2n+4}{2}}}
\int_0^1 \dd u\;
\frac{e^{-u(1-u)\,T}-1}{u^{3/2}\,(1-u)^{1/2}}
\left[\frac{1}{(1-u)^{n}}+\frac{1}{u^{n}}\right]\;.
\end{equation}
The $T$ integral is evaluated by dimensional regularization,
$\int_0^\infty T^{-\alpha}\,e^{-v T}\,\dd T = v^{\alpha-1}\,\Gamma(1-\alpha)$,
and the resulting $u$-integral factorizes into Beta functions. One obtains:
\begin{align}
{\mathcal C}_n(d) \;=\; 
\frac{\Gamma\!\big(\!-\tfrac{d+2n+2}{2}\big)}
     {\Gamma(d+n+2)}
\left[
\Gamma\!\Big(\tfrac{d+2n+1}{2}\Big)\,\Gamma\!\Big(\tfrac{d+3}{2}\Big)
+ \Gamma\!\Big(\tfrac{d+1}{2}\Big)\,\Gamma\!\Big(\tfrac{d+2n+3}{2}\Big)
\right]\;.
\end{align}
The leading (Dirichlet, $n=0$) contribution recovers $\gamma_D$. For $d$ odd,
the Gamma functions are finite for all $n$. 
For instance, setting $d = 2q+1$:
\begin{equation}
\gamma_0^{\rm sub}  \,=\, \frac{(-1)^{q+1}(q!)^2}
{(2\pi)^{q+1}(2q+1)!(2q+3)!!}\;
|k_\parallel|^{2q+3}\;,
\end{equation}
in agreement with Ref.~\cite{FG_Dirichlet}.
The first two corrections, for $d=1$, read:
\begin{equation}
\gamma_1^{\rm sub} \,=\, -\frac{1}{5\pi\,\lambda^2}\;
|k_\parallel|^5 \;, \qquad
\gamma_2^{\rm sub} \,=\, -\frac{8}{21\pi\,\lambda^4}\;
|k_\parallel|^7 \;.
\end{equation}

For even $d$, the factor $\Gamma[-(d+2n+2)/2]$ develops poles. These singularities are treated by dimensional regularization and give rise to logarithmic contributions. As discussed below, these logarithms are the terms that generate the imaginary part of the effective action when $d$ is even. 

\subsection{Imaginary part and pair creation}

To extract the imaginary part of the effective action, which encodes the dissipative effects of the system under consideration, we perform a Wick rotation to Minkowski spacetime. More explicitly, writing $k_\parallel=(k_0,\mathbf{k})$, the Wick rotation is implemented through the continuation $k_0\to -\mathrm{i}(\omega+\mathrm{i}0)$, so that $k_\parallel^2 \rightarrow \mathbf{k}^2-\omega^2-\mathrm{i}0=-p_\parallel^2-\mathrm{i}0$, with $p_\parallel^2 \equiv \omega^2-\mathbf{k}^2$. Together with the corresponding continuation of the effective action, $-\Gamma_2\to \mathrm{i}\Gamma^{(M)}_2$, this prescription fixes the branch of the nonanalytic terms and determines the imaginary part associated with timelike modes.

For odd values of the spatial dimension $d$, the timelike modes of the surface give an imaginary contribution because of the half-integer power in the factor $(-p_\parallel^2)^{\frac{d+2n+2}{2}}$. Setting $d=2q+1$, this can be written as

\begin{align} \label{imaginary_odd}
\mathrm{Im}\big(\Gamma_2^{(M)}\big)&= \frac{1}{2} \, \int\frac{\dd^{2q+1}p_\parallel}{(2\pi)^{2q+1}} \, \Theta(p_\parallel^2) 
\, 
\big|\widetilde{\psi}(p_\parallel)\big|^2\,\sum_{n=0}^{\infty}\zeta(n,q) \,\big(p_\parallel^2\big)^{\frac{2q+2n+3}{2}} \\
\zeta(n,q)&=  \frac{(-1)^n\, 2^{2n}}{\lambda^{2n} (2\pi)^{q+1}} \frac{(2n+1)!!}{(2q+2n+3)!!} \frac{q! (q+n)!}{(2q+n+1)!} 
\end{align}
where $\zeta(0,q)$ corresponds to the dissipation coefficient in the Dirichlet case, which agrees with Ref.~\cite{FG_Dirichlet}. 

For even values of $d$, we use dimensional regularization $d=2q-\epsilon$ and set $\epsilon \to 0$, introducing imaginary contributions through $\log(-\frac{p_\parallel^2}{\mu^2})$, where $\mu^2$ is used for unit consistency. This allows us to write

\begin{align} \label{imaginary_even}
\mathrm{Im}\big(\Gamma_2^{(M)}\big)&= \frac{1}{2} \, \int\frac{\dd^{2q}p_\parallel}{(2\pi)^{2q}} \, \Theta(p_\parallel^2) 
\, 
\big|\widetilde{\psi}(p_\parallel)\big|^2\,\sum_{n=0}^{\infty}\xi(n,q) \,\big(p_\parallel^2\big)^{q+n+1} \\
\xi(n,q)&= \frac{(-1)^n\,\pi^{1-q}(2n+1)!(2q+2n)!(2q)!}{4^{3q+n+1}n!\lambda^{2n}(q+n)!q!(q+n+1)!(2q+n)!}
\end{align}
where $\xi(0,q)$ also gives the correct Dirichlet factor in the mentioned reference.

Several interesting features emerge from these expressions. First, the coefficients alternate in sign with $n$, with even values of $n$ giving positive contributions. This is consistent with the positivity of the leading Dirichlet term, which controls the dominant contribution to the imaginary part of the effective action and, consequently, to the pair-creation probability. The higher-$n$ terms should be understood as finite-coupling corrections to this result: their alternating signs indicate that successive corrections may either enhance or reduce the leading contribution within the regime of validity of the expansion. Second, each higher-$n$ correction has the same momentum-power dependence as the Dirichlet contribution in the shifted dimension $d+2n$. Finally, the factor $\Theta(p_\parallel^2)$ shows that only timelike Fourier components of the surface deformation contribute, as expected for the production of real quanta. In particular, the kinematic threshold for pair creation is the same for all $n$.

\subsection{Exact form factor at finite coupling}\label{sec:continuous}

In this subsection we evaluate \eqref{eq:gamma_def} using \eqref{eq:Ifinal},
obtaining the exact form factor $\gamma(k_\parallel)$ valid for arbitrary
coupling strength $\lambda$.  Substituting the full
propagator~\eqref{eq:kernel}
into~\eqref{eq:Ifinal} and then into~\eqref{eq:gamma_def}, we arrive at one of
the central results of this work:

\begin{align} \label{eq:gamma_complete}
\gamma(k_\parallel)&=-\frac{\lambda^2}{2(4\pi)^{(d+1)/2}} \int_0^\infty \frac{\dd T}{T^{\frac{d+3}{2}}} \int_0^T\dd\tau_1 \int_0^{\tau_1}\dd\tau_2\, e^{k_\parallel^2[\frac{(\tau_1-\tau_2)^2}{T}-(\tau_1-\tau_2)]} \sqrt{T} \times \\ \nonumber    
&\Big[ \frac{1}{2\sqrt{\pi}}\big( \frac{1}{(T-T_b)^{\frac{1}{2}}T_b^{3/2}} +
\frac{1}{(T-T_b)^{3/2}T_b^{\frac{1}{2}}}\big) \\ \nonumber
&-\frac{\lambda}{4}\big(\frac{e^{\frac{\lambda^2}{4}(T-T_b)}}{T_b^{\frac{3}{2}}}\erfc(\frac{\lambda}{2}\sqrt{T-T_b}\;)
 +
\frac{e^{\frac{\lambda^2}{4}(T_b)}}{(T-T_b)^{\frac{3}{2}}}\erfc(\frac{\lambda}{2}\sqrt{T_b}\;)\big)
 \Big]\,.
\end{align}

This function has UV divergences, as expected, and all of them can
be shown to have the form of a polynomial in $k_\parallel^2$. This is indeed
what one could expect since those divergences should correspond to local
counterterms, that could be introduced as part of an action for the surface.
Being then a polynomial, they cannot contribute to the imaginary part of the
effective action, since the latter is always non-analytic in momenta,
and finite.

The integral is regularized using the standard BPHZ prescription, named after Bogoliubov, Parasiuk, Hepp, and Zimmermann, by subtracting a polynomial in $k_\parallel^2$ of degree $N=\lfloor d/2\rfloor$.

Indeed, setting $T_b = \tau_1 - \tau_2$, this subtraction is achieved by
replacing the momentum-dependent exponential in the integrand with its Taylor
remainder up to order $N$:
\begin{equation}\label{eq:bphz}
e^{-k_\parallel^2 \frac{T_b(T-T_b)}{T}} \to e^{-k_\parallel^2
\frac{T_b(T-T_b)}{T}} - \sum_{n=0}^{\lfloor d/2 \rfloor} \frac{(-1)^n}{n!}
\left( k_\parallel^2 \frac{T_b(T-T_b)}{T} \right)^n \;.
\end{equation}
This suppresses both ($T_b \to 0$) and  small-$T$ singularities.
It shifts the proper-time integration scaling from  $T^{-\frac{d}{2}-1}$ to
$T^{\lfloor d/2 \rfloor - \frac{d}{2}}$, securing both UV and IR convergence.
The degree of the subtraction is, as always, {\em sufficient\/} to render the
expression finite.

The subtraction of this polynomial in $k_\parallel^2$ to render the expression
finite can be rephrased as follows: the divergent part is  local in momentum
space and has the form:
\begin{equation}\label{eq:gammadiv}
\gamma_{\rm div}(k_\parallel) \;=\;
\sum_{n=0}^{\lfloor d/2\rfloor}
c_n(d,\lambda)
\left(k_\parallel^2\right)^n \;,
\end{equation}
(where some of the coefficients may be absent).
Hence it does not contribute to the DCE.

Let us now study the finite part of $\gamma(k_\parallel)$, which we denote by
$\gamma_{\rm sub}(k_\parallel)$, obtained by performing the subtraction
above~\footnote{
Note that, since we are interested here in non analytic terms in momentum,
the different renormalization conditions one could have used will not affect
our results.}. We start by decomposing \eqref{eq:gamma_complete} as $\gamma(k_\parallel)=\gamma_1(k_\parallel)+\gamma_2(k_\parallel)$, where $\gamma_1$ collects the free-propagator contribution and $\gamma_2$ contains the $\erfc$ terms.

We begin with $\gamma_1(k_\parallel)$. After integrating over $\tau_1$ and renaming $T_b=\tau$, we obtain

\begin{align} \label{eq:gamma1_def}
    \gamma_1(k_\parallel)&=-\frac{\lambda^2}{2(4\pi)^{(d+1)/2}} \int_0^\infty \frac{\dd T}{T^{\frac{d+2}{2}}} \int_0^T \dd\tau (T-\tau)\, e^{k_\parallel^2[\frac{\tau^2}{T}-\tau]} \\ \nonumber    
& \frac{1}{2\sqrt{\pi}}\big( \frac{1}{(T-\tau)^{\frac{1}{2}}\tau^{3/2}} + \frac{1}{(T-\tau)^{3/2}\tau^{\frac{1}{2}}}\big) \,,
\end{align}
which, after the change of variable $\tau=T \mu$, leaves us with

\begin{align}\label{eq:gamma1_final}
    \gamma_1(k_\parallel)&=-\frac{\lambda^2}{2(4\pi)^{(d+2)/2}} \int_0^\infty \frac{\dd T}{T^{\frac{d+2}{2}}} \int_0^1 \dd\mu \, e^{-T k_\parallel^2\mu(1-\mu)} \\ \nonumber    
&\big( \frac{(1-\mu)^{1/2}}{\mu^{3/2}} + \frac{1}{(1-\mu)^{1/2}\mu^{1/2}}\big) \\ 
&=-\lambda^2 (k_\parallel^2)^{\frac{d}{2}} \frac{\Gamma(-\frac{d}{2}) \Gamma(\frac{d-1}{2}) \Gamma(\frac{d+1}{2})}{2(4\pi)^{\frac{d+2}{2}}\Gamma(d)}\,.
\end{align}

Equation~\eqref{eq:gamma1_final} is well defined for all odd $d>1$, while for even $d$ it must be understood through dimensional regularization, as expected.

For $\gamma_2(k_\parallel)$ one finds

\begin{align} \label{eq:gamma2_def}
    \gamma_2(k_\parallel)&=\frac{\lambda^3}{8(4\pi)^{(d+1)/2}} \int_0^\infty \frac{\dd T}{T^{\frac{d+2}{2}}} \int_0^T \dd\tau (T-\tau)\, e^{k_\parallel^2[\frac{\tau^2}{T}-\tau]} \\ \nonumber    
& \big(\frac{e^{\frac{\lambda^2}{4}(T-\tau)}}{\tau^{\frac{3}{2}}}\erfc(\frac{\lambda}{2}\sqrt{T-\tau}\;) + \frac{e^{\frac{\lambda^2}{4}(\tau)}}{(T-\tau)^{\frac{3}{2}}}\erfc(\frac{\lambda}{2}\sqrt{\tau}\;)\big)\,,
\end{align}
where we use $\erfc(z)=\frac{2}{\sqrt{\pi}}\int_z^\infty e^{-x^2}dx$, and we do the change of variables $\tau=\mu T$, and for the $\erfc$ expressions we use $x=\sqrt{T}y$ and $x=\sqrt{T}z$, leading to

\begin{align}\label{eq:gamma2_intermediate}
\gamma_2(k_\parallel)
&=\frac{\lambda^3}{2(4\pi)^{(d+2)/2}}
\int_0^\infty \frac{\dd T}{T^{\frac{d+2}{2}}}\, T
\int_0^1 \dd\mu \;e^{-k_\parallel^2T\mu(1-\mu)}
\nonumber\\
&\quad\times
\Bigg[
\frac{T(1-\mu)}{T^{\frac{3}{2}}\mu^{\frac{3}{2}}}
e^{\frac{\lambda^2}{4}T(1-\mu)}\sqrt{T}\int_{y_{\text{min}}}^{\infty}e^{-T y^2}\,\dd y
\nonumber\\
&\qquad\qquad
+ \frac{1}{T^{\frac{1}{2}}(1-\mu)^{\frac{1}{2}}}
e^{\frac{\lambda^2}{4}T\mu}\sqrt{T}\int_{z_{\text{min}}}^{\infty}e^{-T z^2}\,\dd z
\Bigg]\,,
\end{align}
where $y_{\text{min}}=\frac{\lambda}{2}\sqrt{1-\mu}$ and $z_{\text{min}}=\frac{\lambda}{2}\sqrt{\mu}$.

Using the symmetry under $\mu \to (1-\mu)$, we obtain

\begin{align}\label{eq:gamma2_reduced}
\gamma_2(k_\parallel) &= \frac{\lambda^3}{2(4\pi)^{(d+2)/2}} \int_0^\infty \frac{\dd T}{T^{\frac{d}{2}}} \int_0^1 \frac{\dd\mu}{\mu^{3/2}} \; e^{-T \left[ k_\parallel^2 \mu(1-\mu) - \frac{\lambda^2}{4}(1-\mu) \right]} \int_{y_{\text{min}}}^{\infty} \dd y \; e^{-T y^2} \nonumber\\
&=\frac{\lambda^3\Gamma(1-\frac{d}{2})}{2(4\pi)^{(d+2)/2}} \int_0^1 \dd\mu\, \mu^{-\frac{3}{2}} \eta(k_\parallel^2, \mu, \lambda)\,,
\end{align}
where we integrated $T$ with analytic continuation, and we define

\begin{align}\label{eq:eta_def} \nonumber
    \eta(k_\parallel^2, \mu, \lambda)&=\int_{y_{\text{min}}}^{\infty}\dd y\, [k_\parallel^2\mu(1-\mu)+y^2-y_{\text{min}}^2]^{\frac{d}{2}-1}\\
    &=\frac{1}{2}\int_0^\infty \dd t\, \big(k_\parallel^2 \mu (1-\mu)+t\big)^{\frac{d}{2}-1} \big(\frac{\lambda^2}{4}(1-\mu)+t\big)^{-\frac{1}{2}} \,,
\end{align}
after the change of variables $t=y^2-y_{\text{min}}^2$.

The integral~\eqref{eq:eta_def} is of the form covered by formula 3.197.1 of Ref.~\cite{GR}:
\begin{equation}\label{eq:GR_formula}
\int_0^\infty x^{\nu-1}(x+a)^{-\mu}(x+b)^{-\rho}\,dx
= a^{-\mu}\,b^{\nu-\rho}\,B(\nu,\mu-\nu+\rho)\;
{}_2F_1\!\big[\mu,\nu;\mu+\rho;1-\tfrac{b}{a}\big]\,,
\end{equation}
valid for $\mathrm{Re}\,\nu > 0$ and $\mathrm{Re}(\mu+\rho-\nu)>0$.

Identifying $x=t$, $a = A \equiv k_\parallel^2\,\mu(1-\mu)$, 
$b = B \equiv \tfrac{\lambda^2}{4}(1-\mu)$, with exponents 
$\nu=1$, $\mu = 1-\tfrac{d}{2}$, $\rho = \tfrac{1}{2}$,
the convergence condition $\mathrm{Re}(\mu+\rho-\nu) = \tfrac{1}{2}-\tfrac{d}{2} > 0$
restricts us to $d<1$; for $d\geq 1$ the result is defined by analytic continuation in $d$, as is standard in dimensional regularization.
With these identifications, and using the simplification
$B(1,\tfrac{1}{2}-\tfrac{d}{2}) = \Gamma(\tfrac{1}{2}-\tfrac{d}{2})/\Gamma(\tfrac{3}{2}-\tfrac{d}{2}) = ({\tfrac{1}{2}-\tfrac{d}{2}})^{-1}$,
we obtain:
\begin{equation}\label{eq:eta_closed}
\eta(k_\parallel^2, \mu, \lambda)
= \frac{\lambda}{2(1-d)}\;
\big[k_\parallel^2\mu(1-\mu)\big]^{\frac{d}{2}-1}
\sqrt{1-\mu}\;\;
{}_2F_1\!\Big(1-\tfrac{d}{2},\,1;\;\tfrac{3}{2}-\tfrac{d}{2};\;
1 - \frac{\lambda^2}{4 k_\parallel^2 \mu}\Big)\,.
\end{equation}

An alternative representation, useful in the regime 
$\lambda \gg |k_\parallel|$, follows from Pfaff's transformation
${}_2F_1(a,b;c;z) = (1-z)^{-a}\,{}_2F_1(a,c-b;c;\tfrac{z}{z-1})$:
\begin{equation}\label{eq:eta_pfaff}
\eta = \frac{\lambda^{d-1}}{2^d\,(1-d)}\;
\frac{(1-\mu)^{\frac{d}{2}-\frac{1}{2}}}{\mu}\;\;
{}_2F_1\!\Big(1-\tfrac{d}{2},\,\tfrac{1}{2}-\tfrac{d}{2};\;\tfrac{3}{2}-\tfrac{d}{2};\;
1 - \frac{4 k_\parallel^2 \mu}{\lambda^2}\Big)\,.
\end{equation}
In this form the argument of the hypergeometric function is 
$1 - 4k_\parallel^2\mu/\lambda^2$, which lies close to $1$ when $\lambda$ is large, making it well suited for expansions in that regime.

Substituting \eqref{eq:eta_closed} into the expression for $\gamma_2$,
we arrive at
\begin{align}\label{eq:gamma2_hypergeometric}
\gamma_2(k_\parallel)
&= \frac{\lambda^4\,\Gamma(1-\frac{d}{2})}{8\,(4\pi)^{\frac{d+2}{2}}}\,
(k_\parallel^2)^{\frac{d}{2}-1}
\int_0^1 \dd\mu\;\mu^{\frac{d}{2}-\frac{5}{2}}\,(1-\mu)^{\frac{d}{2}-\frac{1}{2}}
\nonumber\\
&\quad\times\;
{}_2F_1\!\Big(1-\tfrac{d}{2},\,1;\;\tfrac{3}{2}-\tfrac{d}{2};\;
1-\frac{\lambda^2}{4k_\parallel^2\,\mu}\Big)\,,
\end{align}
where we have used $\tfrac{1}{2}(\tfrac{1}{2}-\tfrac{d}{2})/(1-d) = \tfrac{1}{4}$
to simplify the overall prefactor.

The integral \eqref{eq:gamma2_hypergeometric} requires the subtraction introduced previously to render it UV finite. The apparent poles for odd values of $d$ are artifacts of the decomposition $\gamma=\gamma_1+\gamma_2$, which generates purely real counterterms and therefore does not affect the dissipative contributions. For even values of $d$ (or more precisely, for $d$ treated as a continuous 
parameter via dimensional regularization), both $\gamma_1$ and $\gamma_2$ are 
individually well defined.

\begin{figure}[ht]
\centering
\includegraphics[width=\textwidth]{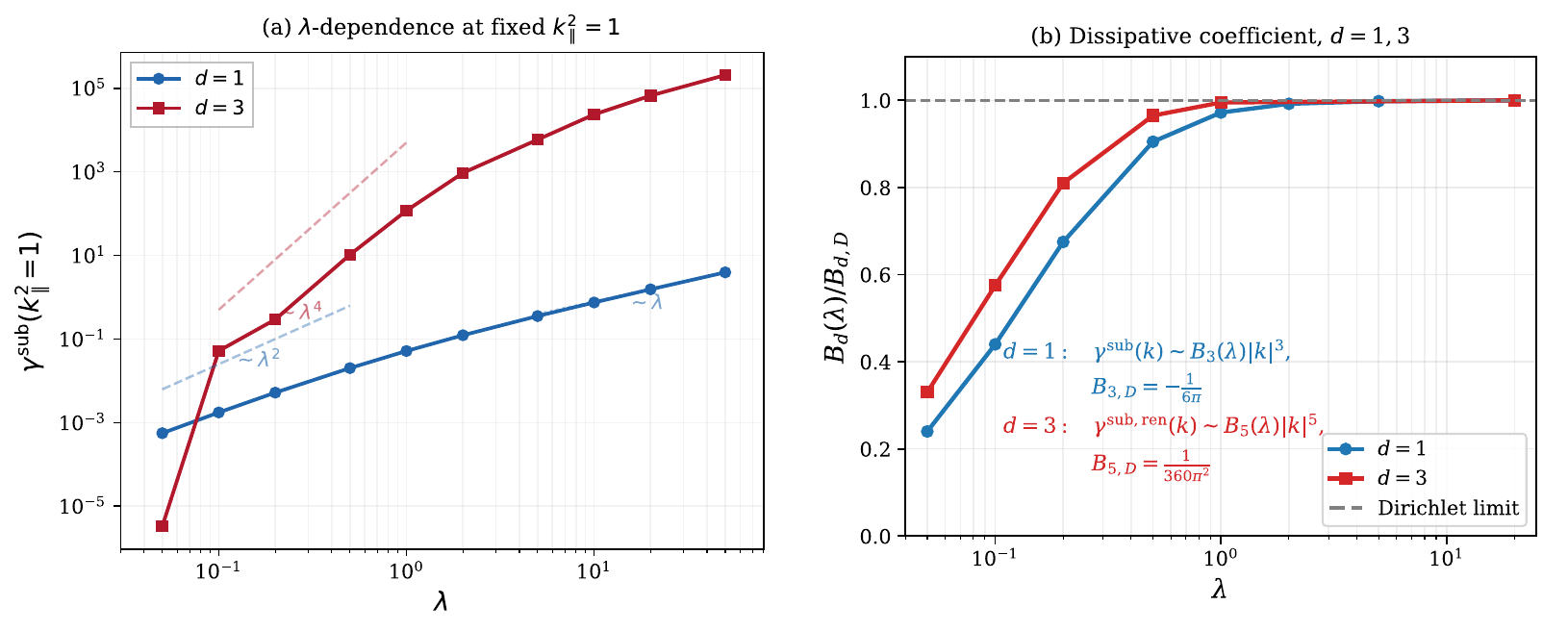}
\caption{%
Subtracted form factor $\gamma^{\rm sub}(k_\parallel) = \gamma(k_\parallel) - \gamma(0)$ 
obtained by numerical evaluation of the full expression~\eqref{eq:gamma_complete}.
\textbf{(a)}~Dependence on the coupling $\lambda$ at fixed $k_\parallel^2 = 1$ 
for $d=1$ (circles) and $d=3$ (squares). 
At weak coupling, $\gamma^{\rm sub} \sim \lambda^2$ for $d=1$ and 
$\gamma^{\rm sub} \sim \lambda^4$ for $d=3$ (dashed guides); 
at strong coupling, the $d=1$ curve crosses over to approximately linear growth, 
reflecting the accumulation of subleading $1/\lambda$ corrections at fixed momentum.
\textbf{(b)}~Ratio of the dissipative coefficient $B(\lambda)$ to its 
Dirichlet-limit value $B_{d, D}$, for $d=1$ and $d=3$.  Here $B(\lambda)$ denotes the 
$|k_\parallel|^m$ coefficient extracted from $\gamma^{\rm sub}$ (for $d=1$, $m=3$, and for $d=3$, $m=5$) by fitting 
the low-momentum expansion $\gamma^{\rm sub} = A(\lambda)\,k_\parallel^2 + \cdots
B(\lambda)\,|k_\parallel|^m + \ldots\;$. As $\lambda \to \infty$, the ratio
$B(\lambda)/B_{d,D} \to 1$, recovering the Dirichlet result~\cite{FG_Dirichlet}
with $B_{1,D} = -1/(6\pi)$ and $B_{3,D}=\frac{1}{360\pi^2}$.}
\label{fig:gamma_interpolation}
\end{figure}

For general $d$ and arbitrary ratio $\zeta \equiv \lambda^2/(4k_\parallel^2)$,
the $\mu$-integral in \eqref{eq:gamma2_hypergeometric} is most efficiently 
evaluated numerically. We have carried out this computation, obtaining the full 
$\lambda$-interpolation of the form factor $\gamma^{\rm sub}$ between the 
weak-coupling ($\lambda \to 0$) and Dirichlet ($\lambda \to \infty$) limits.
The results are displayed in Fig.~\ref{fig:gamma_interpolation}.

\subsection{Higher-order terms}
The third-order contribution to the effective action involves the term
$\psi^3$ from the Taylor expansion of $V$.  After factorization, its 
parallel part requires the worldline average of three $\psi$ insertions:
\begin{equation}
	\int \dd^dx_\parallel\,\langle 
	\psi(x_\parallel(\tau_1))
	\psi(x_\parallel(\tau_2))
	\psi(x_\parallel(\tau_3))
	\rangle_{x_\parallel}\;.
\end{equation}
After performing the $x_\parallel$ integration and the Gaussian worldline 
average, this evaluates to:
\begin{equation}
	\int \frac{\dd^dk_1}{(2\pi)^d}\frac{\dd^dk_2}{(2\pi)^d}\;
	\widetilde{\psi}(k_1)\,\widetilde{\psi}(k_2)\,
	\widetilde{\psi}(-k_1-k_2)\;
	e^{\frac{1}{2}\sum_{i,j}k_i \cdot k_j\,\Delta(\tau_i - \tau_j)}\;.
\end{equation}
However, the third-order terms in the expansion of $e^{-\int V}$
involve an odd number of $v^{(n)}$ insertions in the perpendicular 
sector.  For the $\delta$-potential case, $v(x_d) = \lambda\,\delta(x_d)$, 
the perpendicular path integral is governed by a Hamiltonian that is
even under $x_d \to -x_d$. Therefore, the integrand for each perpendicular 
average contains an odd power of $\delta'(x_d)$ (which is odd under 
reflection), rendering the integral zero. This is not, however, something which 
is restricted to the $\delta$-potential case: we only need an even potential, 
namely $v(-x_d) = v(x_d)$.

More explicitly, the three contributions at third order in $\psi$ are:
(i) a single $v'''$ insertion, (ii) a $v'' \cdot v'$ cross term, and 
(iii) a cubic $v'^3$ term.  In all three cases, the total number of 
derivatives of $v$ acting on the perpendicular coordinate is odd.  
Since $K(t;\,x,y)$ depends on $|x|$ and $|y|$ (and is therefore even under 
$x_d\to -x_d$), and since the derivatives of odd order of the potential are odd 
functions, each perpendicular integral vanishes.  This argument extends to all 
odd orders in $\psi$:
\begin{equation}
	\Gamma_{2n+1}(\psi) \,=\, 0\;, \qquad n = 0, 1, 2, \ldots
\end{equation}
for any even potential.

The fourth-order contribution $\Gamma_4(\psi)$ receives five types of terms, 
arising from the expansion of $\langle e^{-\int_0^T \dd\tau\,V}\rangle$ 
to fourth order in $\psi$.
We write them schematically as:
\begin{equation}
	\Gamma_4 \;=\; \Gamma_{4,1} + \Gamma_{4,2} + \Gamma_{4,3} + \Gamma_{4,4}
	+ \Gamma_{4,5}\;,
\end{equation}
where the subscripts indicate the partition of four $\psi$ insertions into 
groups associated with different $v^{(n)}$ vertices:

\noindent (a) $\Gamma_{4,1}$: a single $v^{(4)}$ insertion ($n=4$ in the 
Taylor expansion, one proper-time variable);

\noindent (b) $\Gamma_{4,2}$: a $v''' \cdot v'$ cross term ($n=3$ and $n=1$, 
two proper-time variables);

\noindent (c) $\Gamma_{4,3}$: a $(v'')^2$ term ($n=2$ twice, two 
proper-time variables);

\noindent (d) $\Gamma_{4,4}$: a $v'' \cdot (v')^2$ term ($n=2$ and two 
$n=1$ insertions, three proper-time variables);

\noindent (e) $\Gamma_{4,5}$: a $(v')^4$ term ($n=1$ four times, four 
proper-time variables).

Each contribution factorizes into a parallel part (involving products of 
$\psi$ in Fourier space) and a perpendicular part (a worldline average 
of derivatives of $v$ with the full $e^{-\int v}$ insertion).

\section{Two surfaces}\label{sec:twosurfaces}

In this section we analyze the case of two surfaces, one of them flat, 
located at $x_d=a$ (we leave the sign of $a$ free), and one curved 
surface located at $x_d=\psi(x_\parallel)$. This setup leaves us 
with $V(x)=\lambda_1 \,\delta(x_d-\psi(x_\parallel)) + \lambda_2 
\,\delta(x_d-a)$.  The two-surface dynamical Casimir effect has been 
studied in a functional-determinant framework for imperfect mirrors with 
nonlocal couplings~\cite{FLM_2007,FLM_2011}; here we show how the 
worldline approach naturally incorporates the second surface through 
the two-$\delta$ propagator.

Following the procedure of Sect.~\ref{sec:perturbative}, we focus on 
the first term in the power expansion that can generate dissipative 
effects, namely
\begin{align} \label{eq:gamma22_two_surfaces}
\Gamma_{2,2}(\psi)
&= - \frac{\lambda_1^2}{4(4\pi)^{\frac{d+1}{2}}}
\int_{0^+}^\infty \frac{\dd T}{T^{\frac{d+3}{2}}}
\int_0^T\dd\tau_1 \int_0^{\tau_1}\dd\tau_2
\nonumber\\
&\quad\times
\Bigg[
\int \dd^dx_\parallel \,
\langle \psi(x_\parallel(\tau_1)) \psi(x_\parallel(\tau_2)) \rangle_{x_\parallel}
\nonumber\\
&\qquad\times
\int \dd x_d \,
\langle \delta'(x_d(\tau_1)) \delta'(x_d(\tau_2))
e^{-\int_0^T \dd\tau\,V(x_d(\tau))}\rangle_{x_d}
\Bigg] \,,
\end{align}
where $V(x_d(\tau))=\lambda_1 \,\delta(x_d) + \lambda_2 \,\delta(x_d-a)$. 

The difference between \eqref{eq:gamma22_two_surfaces} and the 
single-surface case~\eqref{eq:gamma22_one_surface} is captured entirely 
by the difference between the Euclidean propagator for a particle in the 
presence of one $\delta$-potential at the origin, 
eq.~\eqref{eq:kernel}, and the propagator $K^{(\delta_2)}$ 
corresponding to two $\delta$-potentials, one at the origin and one 
at $x_d=a$.  The latter may be represented via its resolvent as follows:
\begin{equation} \label{eq:kernel_green_relation}
K^{(\delta_2)}(t,x,y) = -i\int_{-\infty}^{\infty}\frac{\dd E}{2\pi}\, 
e^{-E\,t}\; G^{(\delta_2)}(E,x,y)\;,
\end{equation}
where, following~\cite{GroscheSteiner}, the two-$\delta$ Green function 
is expressed in terms of the single-$\delta$ one:
\begin{equation}\label{eq:Gdelta2}
G^{(\delta_2)}(E,x,y)=G^{(\delta_1)}(E,x,y) 
- \frac{G^{(\delta_1)}(E,x,a)\;G^{(\delta_1)}(E,a,y)}
{G^{(\delta_1)}(E,a,a) - \frac{1}{\lambda_2}}\;,
\end{equation}
with
\begin{equation}\label{eq:Gdelta1}
G^{(\delta_1)}(E,x,y)= \frac{e^{-\sqrt{-E}\,|x-y|}}{2\sqrt{-E}} 
+ \frac{\lambda_1\, e^{-\sqrt{-E}\,(|x|+|y|)}}
{2\sqrt{-E}\bigl(2\sqrt{-E} - \lambda_1\bigr)} \;.
\end{equation}
The inverse Laplace transform of $G^{(\delta_1)}$ 
via~\eqref{eq:kernel_green_relation} reproduces the single-$\delta$ kernel 
$K$ of eq.~\eqref{eq:kernel} (with $\lambda = \lambda_1$).

We decompose the two-$\delta$ propagator as
\begin{equation} \label{eq:K_delta2_def}
K^{(\delta_2)}(t,x,y) \;=\; K(t,x,y) + \delta K(t,x,y)\;,
\end{equation}
where the correction $\delta K$ due to the second wall is
\begin{align} \label{eq:deltaK_spectral}
\delta K(t,x,y) &= -i\int_{-\infty}^{\infty}\frac{\dd E}{2\pi}\, 
e^{-E\,t}\;
\frac{\lambda_2}{2\sqrt{-E}\,(2\sqrt{-E} - \lambda_1)} 
\;\frac{A(E,x)\,A(E,y)}{\mathcal D(E)}\;,
\end{align}
with
\begin{align}
A(E,z) &= (2\sqrt{-E}-\lambda_1)\, e^{-\sqrt{-E}\,|z-a|}
+ \lambda_1\, e^{-\sqrt{-E}\,(|z|+|a|)} \;, \label{eq:defA}\\[3pt]
\mathcal D(E) &= (2\sqrt{-E}-\lambda_1)(2\sqrt{-E}-\lambda_2) 
- \lambda_1\,\lambda_2\;e^{-2\sqrt{-E}\,|a|} \;.
\label{eq:defD}
\end{align}

Since the parallel sector of the worldline average is identical to the 
single-surface case, the full result adopts the same quadratic form:
\begin{equation}\label{eq:Gamma22_chi}
\Gamma_{2,2}(\psi)=\frac{1}{2} 
\int\frac{\dd^dk_\parallel}{(2\pi)^d}\; 
\chi(k_\parallel)\; |\widetilde{\psi}(k_\parallel)|^2\,,
\end{equation}
with 
\begin{equation}\label{eq:chi_def}
\chi(k_\parallel)=-\frac{\lambda_1^2}{2(4\pi)^{d/2}} 
\int_0^\infty \frac{\dd T}{T^{(d+2)/2}} \int_0^T\dd\tau_1 
\int_0^{\tau_1}\dd\tau_2 \;e^{k_\parallel^2 \Delta(\tau_1-\tau_2)}\; 
I^{(\delta_2)}(T,\tau_1,\tau_2)\,.
\end{equation}
Here $I^{(\delta_2)}$ is obtained by replacing 
$K \to K^{(\delta_2)}$ in the four-term structure of 
eq.~\eqref{eq:I_reduced}.  The factor 
$(4\pi T)^{1/2}\lambda^2$ that appears in the single-surface definition 
of $I$ (eq.~\eqref{eq:I_reduced}) is absorbed into the prefactor and 
the power of $T$ in eq.~\eqref{eq:chi_def}, exactly as in the passage 
from eq.~\eqref{eq:I_reduced} to eq.~\eqref{eq:gamma_def}.  Keeping the 
same definitions $T_b = \tau_1 - \tau_2$ and $S = T - T_b$ as before, 
the result is:
\begin{align} \label{eq:I_delta2_def}
I^{(\delta_2)} &= 
\partial_1\partial_2 K^{(\delta_2)}_b(0,0)\;K^{(\delta_2)}_S(0,0)
\;+\;
K^{(\delta_2)}_b(0,0)\;\partial_1\partial_2 K^{(\delta_2)}_S(0,0) 
\nonumber\\
&\;+\;
\partial_1 K^{(\delta_2)}_b(0,0)\;\partial_2 K^{(\delta_2)}_S(0,0)
\;+\;
\partial_2 K^{(\delta_2)}_b(0,0)\;\partial_1 K^{(\delta_2)}_S(0,0)\;,
\end{align}
where $K^{(\delta_2)}_b \equiv K^{(\delta_2)}(T_b;\,\cdot\,,\cdot)$ and
similarly for $K^{(\delta_2)}_S$.

Introducing~\eqref{eq:K_delta2_def} into~\eqref{eq:I_delta2_def}, we write 
$I^{(\delta_2)}=I^{(\delta_1)} + \delta I$, where $I^{(\delta_1)}$ is 
the single-surface contribution (setting $\lambda_1=\lambda$), and 
$\delta I$ collects all terms involving at least one factor of $\delta K$.
Setting $\chi=\gamma+\delta \gamma$, we need only compute 
$\delta \gamma$:
\begin{equation}\label{eq:dgamma_def}
\delta \gamma(k_\parallel)=-\frac{\lambda_1^2}{2(4\pi)^{d/2}} 
\int_0^\infty \frac{\dd T}{T^{(d+2)/2}} \int_0^T\dd\tau_1 
\int_0^{\tau_1}\dd\tau_2 \;e^{k_\parallel^2 \Delta(\tau_1-\tau_2)}\; 
\delta I(T,\tau_1,\tau_2)\,.
\end{equation}

We organize $\delta I= \delta I_1 + \delta I_2$ by the order in 
$\delta K$.  The terms linear in $\delta K$ are:
\begin{align}\label{eq:dI1_def}
\delta I_1 &= \partial_1\partial_2 \delta K(T_b;0,0)\;K(S;0,0)
\;+\;
\partial_1\partial_2 K(T_b;0,0)\;\delta K(S;0,0) \nonumber\\
&\;+\;
\delta K(T_b;0,0)\;\partial_1\partial_2 K(S;0,0)
\;+\;
K(T_b;0,0)\;\partial_1\partial_2 \delta K(S;0,0) \,,
\end{align}
where we used the single-$\delta$ identity 
$\partial_i K(t;\,0,0)=0$.  The terms quadratic in $\delta K$ are:
\begin{align}\label{eq:dI2_def}
\delta I_2 &= \partial_1\partial_2 \delta K(T_b;0,0)\;\delta K(S;0,0)
\;+\;
\delta K(T_b;0,0)\;\partial_1\partial_2 \delta K(S;0,0) \nonumber\\
&\;+\;
2\;\partial_1 \delta K(T_b;0,0)\;\partial_1 \delta K(S;0,0)\;,
\end{align}
where the factor of $2$ in the last term arises because $\delta K$ 
is symmetric in its spatial arguments, so 
$\partial_1\delta K(t;\,0,0) = \partial_2\delta K(t;\,0,0)$, and the 
two cross terms in~\eqref{eq:I_delta2_def} are equal.

To proceed, we need explicit expressions for $\delta K$ and its 
derivatives evaluated at $x = y = 0$.  The key simplification comes 
from the function $A(E,z)$, which at $z = 0$ reduces to
\begin{equation}\label{eq:A0}
A(E,0) \;=\; (2p - \lambda_1)\,e^{-p|a|} + \lambda_1\,e^{-p|a|} 
\;=\; 2p\,e^{-p|a|}\;,
\end{equation}
where $p \equiv \sqrt{-E}$.  For the derivative with respect to $z$, 
we note that $\partial_z|z-a|\big|_{z=0} = -\sgn(a)$ (well defined 
for $a \neq 0$), while $\partial_z|z|\big|_{z=0} = 0$ in the 
symmetric-derivative sense appropriate to $\int f\,\delta'\,\dd x = -f'(0)$. This gives
\begin{equation}\label{eq:A1}
A'(E,0) \;\equiv\; \partial_z A(E,z)\big|_{z=0}
\;=\; p\,(2p - \lambda_1)\,\sgn(a)\;e^{-p|a|}\;.
\end{equation}
Since $\delta K(t;\,x,y)$ factorizes as $A(E,x)\,A(E,y)$ (times scalar 
functions of $E$), the derivatives of $\delta K$ at the origin are 
determined by $A(E,0)$ and $A'(E,0)$ alone.  Substituting 
\eqref{eq:A0}--\eqref{eq:A1} into the spectral 
representation~\eqref{eq:deltaK_spectral} and writing 
$\widetilde{\mathcal D}(p) \equiv \mathcal D(-p^2)$, one obtains the three 
independent building blocks:
\begin{align}
\delta K(t;\,0,0) &= \frac{1}{\pi}\int_0^\infty \!\dd p\;
e^{-p^2 t}\;\frac{2\lambda_2\,p\;e^{-2p|a|}}{(2p-\lambda_1)\,
\widetilde{\mathcal D}(p)} \;, \label{eq:dK00}\\[4pt]
\partial_i\,\delta K(t;\,0,0) &= \frac{\sgn(a)}{\pi}
\int_0^\infty \!\dd p\;e^{-p^2 t}\;
\frac{\lambda_2\,p\;e^{-2p|a|}}{\widetilde{\mathcal D}(p)} \;,
\label{eq:d1dK00}\\[4pt]
\partial_1\partial_2\,\delta K(t;\,0,0) &= \frac{1}{\pi}
\int_0^\infty \!\dd p\;e^{-p^2 t}\;
\frac{\lambda_2\,p\,(2p-\lambda_1)\;e^{-2p|a|}}
{2\,\widetilde{\mathcal D}(p)} \;, \label{eq:d12dK00}
\end{align}
with
\begin{equation}\label{eq:Dtilde}
\widetilde{\mathcal D}(p) = (2p - \lambda_1)(2p - \lambda_2) 
- \lambda_1\,\lambda_2\;e^{-2p|a|}\;.
\end{equation}
The integrals above are spectral (inverse-Laplace) representations, obtained
by collapsing the energy contour onto the continuum branch cut $E\le 0$
(real $p=\sqrt{-E}$). Two remarks regarding their validity are in order. First,
in the regime of interest here-repulsive couplings $\lambda_1,\lambda_2>0$,
and in particular the Dirichlet limit $\lambda_i\to\infty$-the resolvent
$\widetilde{\mathcal D}^{-1}$ has no poles off the cut, so the representations
are exact as written. Second, bound states can occur only when at least one
coupling is attractive; each such state sits at a fixed $p=p_b$ with energy
$E_b=-p_b^2<0$ (a zero of $\widetilde{\mathcal D}$), and adds to the kernels a
discrete contribution $\propto e^{-E_b t}=e^{p_b^2 t}$. Being entire (hence
analytic) in the proper time $t$, such a term integrates to a purely
analytic, {\em local\/} dependence on $k_\parallel^2$; it does not affect the
large-$p$ (UV, $T\to0$) behavior of $\delta\gamma$, and, being local, it
drops out of the imaginary part and is therefore immaterial for the DCE. We
may consequently ignore these terms throughout.

A useful identity is
\begin{equation}\label{eq:d12K_free}
\partial_1\partial_2\,K(t;\,0,0) \;=\; \frac{1}{4\sqrt{\pi}\,t^{3/2}}
\;,
\end{equation}
valid for all values of $\lambda_1$.  This follows from the fact that, 
at $x = y = 0$, the symmetric derivative 
$\partial_x|x|\big|_{x=0} = 0$ kills the $\lambda$-dependent 
correction in~\eqref{eq:kernel}, leaving only the free-propagator 
contribution.


Substituting the building blocks into $\delta I_1$ and using $\partial_i 
K(t;\,0,0) = 0$ together with~\eqref{eq:d12K_free}, the four terms 
become
\begin{align}\label{eq:dI1explicit}
\delta I_1 &= \partial_1\partial_2\,\delta K(T_b)\;K(S)
\;+\; \frac{\delta K(S)}{4\sqrt\pi\,T_b^{3/2}}
\;+\; \frac{\delta K(T_b)}{4\sqrt\pi\,S^{3/2}}
\;+\; K(T_b)\;\partial_1\partial_2\,\delta K(S)\;,
\end{align}
where we have suppressed the arguments $(0,0)$ on all kernels.

For $\delta I_2$, which is quadratic in $\delta K$, all three terms 
in~\eqref{eq:dI2_def} are retained as written.  Since every factor of 
$\delta K$ or its derivatives carries a suppression factor $e^{-2p|a|}$ 
from the spectral integrals, $\delta I_2 \sim e^{-4p|a|}$ in the 
large-$|a|$ regime, while $\delta I_1 \sim e^{-2p|a|}$.  The leading 
correction to $\delta\gamma$ thus comes entirely from $\delta I_1$.

When $|a|$ is large compared to the typical worldline scale 
$\sim 1/\sqrt{k_\parallel^2}$, the exponential factor $e^{-2p|a|}$ 
in the spectral integrals suppresses $\delta K$ and its derivatives.  At 
leading order, we expand $\widetilde{\mathcal D}^{-1}$ by neglecting the 
exponential term in~\eqref{eq:Dtilde}:
\begin{equation}\label{eq:DtildeExpansion}
\frac{1}{\widetilde{\mathcal D}(p)} \;\approx\; 
\frac{1}{(2p-\lambda_1)(2p-\lambda_2)}
\sum_{n=0}^\infty \left[
\frac{\lambda_1\lambda_2\,e^{-2p|a|}}{(2p-\lambda_1)(2p-\lambda_2)}
\right]^n.
\end{equation}
At leading ($n=0$) order, the mixed 
derivative~\eqref{eq:d12dK00} simplifies by a cancellation of 
$(2p - \lambda_1)$:
\begin{equation}\label{eq:d12dK_LO}
\partial_1\partial_2\,\delta K(t;\,0,0) \;\approx\;
\frac{1}{\pi}\int_0^\infty\!\dd p\;e^{-p^2 t}\;
\frac{\lambda_2\,p\;e^{-2p|a|}}{2\,(2p - \lambda_2)}\;,
\end{equation}
while $\delta K$ and $\partial_i\delta K$ retain both couplings:
\begin{align}
\delta K(t;\,0,0) &\approx \frac{1}{\pi}
\int_0^\infty\!\dd p\;
e^{-p^2 t}\;\frac{2\lambda_2\,p\;e^{-2p|a|}}
{(2p-\lambda_1)^2(2p-\lambda_2)}\;, \label{eq:dK_LO}\\[4pt]
\partial_i\,\delta K(t;\,0,0) &\approx
\frac{\sgn(a)}{\pi}\int_0^\infty\!\dd p\;
e^{-p^2 t}\;\frac{\lambda_2\,p\;e^{-2p|a|}}
{(2p-\lambda_1)(2p-\lambda_2)}\;. \label{eq:d1dK_LO}
\end{align}

The correction to the form factor at this order is
\begin{align}\label{eq:dgamma_LO}
\delta\gamma(k_\parallel) &=
-\frac{\lambda_1^2}{2(4\pi)^{d/2}}
\int_0^\infty\!\frac{\dd T}{T^{(d+2)/2}}
\int_0^T\!\dd T_b\,(T - T_b)\;
e^{\Delta(T_b)\,k_\parallel^2}\;
\delta I_1(T_b,\,T - T_b)
\nonumber\\
&\quad+\; O(e^{-4p|a|})\;,
\end{align}
where $\delta I_1$ is given by~\eqref{eq:dI1explicit} with the 
leading-order spectral functions.

The physical content of~\eqref{eq:dgamma_LO} is that the second surface 
modifies the dissipative form factor through an interference between 
the single-wall propagator $K(t;\,0,0)$ and the wall-to-wall correction 
$\delta K(t;\,0,0)$.  This interference is exponentially suppressed 
for large wall separations, reflecting the fact that the worldline must 
traverse the distance $|a|$ to ``feel'' the second surface, at a cost 
$\sim e^{-2p|a|}$ in the spectral integral.

\subsection{Dirichlet limit}\label{sec:Dir_limit}

In the limit $\lambda_1, \lambda_2 \to \infty$ at fixed $|a|$, the 
building blocks exhibit a hierarchy:
\begin{align}\label{eq:Dir_hierarchy}
K(t;\,0,0) &\sim \frac{1}{\lambda_1^2\sqrt{\pi}\,t^{3/2}}\;, &
\delta K(t;\,0,0) &\sim O(1/\lambda_1^2)\;, \nonumber\\[2pt]
\partial_i\,\delta K(t;\,0,0) &\sim O(1/\lambda_1)\;, &
\partial_1\partial_2\,\delta K(t;\,0,0) &\to
\partial_1\partial_2\,\delta K^{\mathrm{Dir}}(t)\;,
\end{align}
where only the mixed derivative remains $O(1)$.  Its limiting form is
\begin{equation}\label{eq:d12dK_Dirichlet}
\partial_1\partial_2\,\delta K^{\mathrm{Dir}}(t) \;=\;
-\frac{1}{2\pi}\int_0^\infty\!\dd p\;e^{-p^2 t}\;
\frac{p\,e^{-2p|a|}}{1 - e^{-2p|a|}} \;.
\end{equation}
The integrand admits a geometric-series expansion
\begin{equation}\label{eq:image_sum}
\frac{p\,e^{-2p|a|}}{1 - e^{-2p|a|}} 
\;=\; \sum_{n=1}^{\infty} p\,e^{-2np|a|}\;,
\end{equation}
interpretable as a sum over image charges at distances $2n|a|$.  This is 
the standard method-of-images result for the correction to the 
Dirichlet propagator due to a second 
wall~\cite{FLM_2011,GroscheSteiner}, providing an independent 
verification of the two-surface formalism.

A further simplification occurs in $\delta I_1$: in the Dirichlet limit, 
$\delta K(t;\,0,0)$ is proportional to 
$\partial_1\partial_2\,\delta K(t;\,0,0)$ with coefficient 
$4/\lambda_1^2$.  Substituting into~\eqref{eq:dI1explicit} and 
factoring out $1/\lambda_1^2$, each of the four terms contributes with 
the same $1/t^{3/2}$ structure from either $K$ or $\delta K$, yielding
\begin{equation}\label{eq:dI1Dir}
\delta I_1^{\,\mathrm{Dir}} \;=\;
\frac{2}{\lambda_1^2\sqrt\pi}\left[
\frac{\partial_1\partial_2\,\delta K^{\mathrm{Dir}}(T_b)}{S^{3/2}}
\;+\; \frac{\partial_1\partial_2\,\delta K^{\mathrm{Dir}}(S)}
{T_b^{3/2}}
\right].
\end{equation}
After multiplication by $\lambda_1^2$ in the prefactor 
of~\eqref{eq:dgamma_def}, the coupling cancels and the result is
\begin{align}\label{eq:dgammaDir}
\delta\gamma^{\,\mathrm{Dir}}(k_\parallel) &=
-\frac{1}{(4\pi)^{d/2}\sqrt\pi}
\int_0^\infty\!\frac{\dd T}{T^{(d+2)/2}}
\int_0^T\!\dd T_b\,(T - T_b)\;
e^{\Delta(T_b)\,k_\parallel^2}
\nonumber\\
&\qquad\times
\left[
\frac{\partial_1\partial_2\,\delta K^{\mathrm{Dir}}(T_b)}{S^{3/2}}
+ \frac{\partial_1\partial_2\,\delta K^{\mathrm{Dir}}(S)}{T_b^{3/2}}
\right].
\end{align}
We have verified numerically that the exact $\delta I_1$ (with finite 
$\lambda_1 = \lambda_2 = \lambda$) converges to~\eqref{eq:dI1Dir} as 
$\lambda \to \infty$, with the ratio approaching unity.  

To obtain a more explicit expression, we substitute the image-charge 
expansion~\eqref{eq:image_sum} into~\eqref{eq:dgammaDir} and rescale 
$T_b = u\,T$.  The proper-time integral over $T$ is then evaluated by 
dimensional regularization (exactly as in the single-surface 
derivation of Sect.~\ref{sec:perturbative}), with the $k_\parallel = 0$ subtraction handled automatically by analytic continuation. The result is
\begin{multline}\label{eq:dgammaDir_explicit}
\delta\gamma^{\,\mathrm{Dir,sub}}(k_\parallel)
=
\frac{\Gamma(-\tfrac{d}{2})}{2\pi^{3/2}(4\pi)^{d/2}}
\sum_{n=1}^{\infty}\int_0^\infty\!\dd p\;p\,e^{-2np|a|}
\int_0^1\!\dd u\,
(1-u)^{-1/2}
\\
\times\Bigl[\bigl(p^2 u\bigr)^{d/2}
- \bigl(u(1-u)\,k_\parallel^2 + p^2 u\bigr)^{d/2}\Bigr]
\\
+
\frac{\Gamma(-\tfrac{d}{2})}{2\pi^{3/2}(4\pi)^{d/2}}
\sum_{n=1}^{\infty}\int_0^\infty\!\dd p\;p\,e^{-2np|a|}
\int_0^1\!\dd u\,
(1-u)\,u^{-3/2}
\\
\times\Bigl[\bigl(p^2(1-u)\bigr)^{d/2}
- \bigl(u(1-u)\,k_\parallel^2 + p^2(1-u)\bigr)^{d/2}\Bigr].
\end{multline}
For odd $d = 2q+1$, $\Gamma(-d/2)$ is finite and the $u$-integrals 
can be evaluated in closed form.  For $d=1$, both integrals reduce 
to logarithms and inverse tangents, yielding:
\begin{align}\label{eq:dgammaDir_d1}
\delta\gamma^{\,\mathrm{Dir,sub}}\big|_{d=1}
&=
-\frac{1}{2\pi^2}\sum_{n=1}^{\infty}\int_0^\infty\!\dd p\;
p\,e^{-2np|a|}
\nonumber\\
&\quad\times
\left[
\frac{2|k_\parallel|}{p}\,\arctan\frac{|k_\parallel|}{p}
- \log\!\left(1+\frac{k_\parallel^2}{p^2}\right)
\right].
\end{align}
The $p$-integral in~\eqref{eq:dgammaDir_d1} can be expressed in terms 
of exponential integrals; in practice it is most efficiently 
evaluated by direct quadrature.  For $|k_\parallel| \ll 1/|a|$ the 
bracket behaves as $\sim k_\parallel^2/p^2$, and for 
$|k_\parallel| \gg 1/|a|$ as 
$\sim \pi|k_\parallel|/p$, interpolating between a 
$k_\parallel^2$ (non-dissipative) and a 
$|k_\parallel|^3$ (dissipative) regime.  
In particular, at large separation, the leading image ($n=1$) dominates 
and the dissipative piece scales as 
$|k_\parallel|^3\,e^{-2|k_\parallel||a|}$, 
exponentially suppressed relative to the single-surface result 
$\gamma_D = -|k_\parallel|^3/(6\pi)$.  This, together with the 
exact agreement of the single-surface Dirichlet limit 
with the direct calculation of 
Ref.~\cite{FG_Dirichlet}, provides an independent verification of the 
two-surface formalism.

The Dirichlet-limit correction~\eqref{eq:dgammaDir_explicit} has a 
structure parallel to the single-surface result: comparing with the 
Dirichlet form factor $\gamma_D$ from~\eqref{eq:gammaD_closed}, the factor 
$1/(4\sqrt{\pi}\,t^{3/2})$ associated with the single-wall Dirichlet 
propagator is replaced by $\partial_1\partial_2\,\delta K^{\mathrm{Dir}}(t)$, 
which encodes the effect of the second wall through the image-charge 
sum~\eqref{eq:image_sum}.  For large $|a|$, 
$\delta\gamma^{\,\mathrm{Dir}}$ is exponentially suppressed, recovering 
the single-surface Dirichlet result.  In the opposite limit 
$|a| \to 0$, the image sum diverges and the perturbative expansion in 
$\delta K$ breaks down, signaling the coalescence of the two surfaces.

\section{Conclusions}\label{sec:conclusions}

We have developed a worldline approach to the Dynamical Casimir Effect for a 
real scalar field coupled to a time-dependent surface modeled by a background 
potential. The key advantage of this formulation is the natural factorization 
of the worldline path integral into parallel and perpendicular sectors, which 
reduces the computation of the effective action to one-dimensional quantum 
mechanical problems.

For the case of a $\delta$-function potential of strength $\lambda$, we have 
obtained an exact expression for the dissipative form factor 
$\gamma(k_\parallel)$, valid for arbitrary coupling strength.  
In the strong-coupling limit $\lambda \to \infty$, we recover the Dirichlet 
result of Ref.~\cite{FG_Dirichlet}, and derive the systematic corrections in 
inverse powers of $\lambda$: the $n$-th order contribution scales as 
$|k_\parallel|^{d+2n+2}/\lambda^{2n}$, with a closed-form coefficient 
$\mathcal{C}_n(d)$ expressed through Gamma functions.

The imaginary part of the effective action was computed for arbitrary values of $d$, including corrections to the coupling strength up to an arbitrary order $n$ in $1/\lambda$. At leading order, the imaginary part is positive, as expected, whereas the signs of the higher-order contributions alternate as $(-1)^n$, for both odd and even spatial dimensions $d$. The momentum-power dependence of the $n$-th order correction matches that of the Dirichlet case in shifted dimension $d+2n$. 

The full $\lambda$-dependent form factor has been obtained in closed form via 
hypergeometric functions, by splitting $\gamma$ into a free-propagator part 
$\gamma_1$ and a $\delta$-potential correction $\gamma_2$. The latter is 
expressed through a ${}_2F_1$ function whose argument interpolates continuously 
between the weak-coupling and Dirichlet regimes. We have verified the 
cancellation of spurious poles between $\gamma_1$ and $\gamma_2$ at odd 
spatial dimensions.

Using a parity argument based on the evenness of the perpendicular Hamiltonian 
under $x_d \to -x_d$, we have shown that all odd-order contributions in 
$\psi$ vanish identically for any even potential, and have classified the five 
distinct types of terms arising at fourth order.

The formalism has been extended to a two-surface configuration, where the 
second surface introduces corrections to the propagator that can be treated 
perturbatively through the Green's function of the two-$\delta$ 
potential system.  In the Dirichlet limit, the correction to the 
form factor reduces to an image-charge sum, recovering the standard 
method-of-images structure and connecting with the functional-determinant 
results of Refs.~\cite{FLM_2007,FLM_2011}.

Several directions remain open. The explicit computation of the fourth-order term $\Gamma_4(\psi)$, together with the Bern-Kosower representation of the resulting amplitude, would provide a diagrammatic interpretation analogous to that of the string-inspired formalism and will be addressed elsewhere. Extensions to the electromagnetic field and to finite temperature are also natural next steps within this framework.

\section*{Acknowledgments}
This work was supported by CONICET, and UNCuyo.


\begin{thebibliography}{99}

\bibitem{DCE_Review} V.~V.~Dodonov, ``Current status of the dynamical 
		Casimir effect,'' Phys.\ Scr.\ \textbf{82}, 038105 (2010).
		\href{https://doi.org/10.1088/0031-8949/82/03/038105}%
		{\texttt{doi:10.1088/0031-8949/82/03/038105}}.

\bibitem{Moore} G.~T.~Moore, ``Quantum theory of the electromagnetic 
		field in a variable-length one-dimensional cavity,''
		J.\ Math.\ Phys.\ \textbf{11}, 2679 (1970).
		\href{https://doi.org/10.1063/1.1665432}%
		{\texttt{doi:10.1063/1.1665432}}.

\bibitem{Davies_Fulling} P.~C.~W.~Davies and S.~A.~Fulling,
		``Radiation from a moving mirror in two dimensional 
		space-time: conformal anomaly,''
		Proc.\ R.\ Soc.\ Lond.\ A \textbf{348}, 393 (1976).
		\href{https://doi.org/10.1098/rspa.1976.0045}%
		{\texttt{doi:10.1098/rspa.1976.0045}}.

\bibitem{Gies_WL} H.~Gies, K.~Langfeld and L.~Moyaerts, ``Casimir effect 
		on the worldline,'' JHEP \textbf{06} (2003) 018
		[arXiv:hep-th/0303264].
		\href{https://doi.org/10.1088/1126-6708/2003/06/018}%
		{\texttt{doi:10.1088/1126-6708/2003/06/018}}.

\bibitem{Worldline_Review} C.~Schubert, ``Perturbative quantum field 
		theory in the string-inspired formalism,'' 
		Phys.\ Rep.\ \textbf{355}, 73 (2001)
		[arXiv:hep-th/0101036].
		\href{https://doi.org/10.1016/S0370-1573(01)00013-8}%
		{\texttt{doi:10.1016/S0370-1573(01)00013-8}}.

\bibitem{GroscheSteiner} C.~Grosche and F.~Steiner,
		\textit{Handbook of Feynman Path Integrals},
		Springer Tracts in Modern Physics \textbf{145} (1998).
		\href{https://doi.org/10.1007/BFb0109520}%
		{\texttt{doi:10.1007/BFb0109520}}.

\bibitem{FG_Dirichlet} C.~D.~Fosco and B.~C.~Guntsche,
		``Quantum dissipative effects for a real scalar field 
		coupled to a time-dependent Dirichlet surface in $d+1$ 
		dimensions,'' Phys.\ Rev.\ D \textbf{109}, 065023 (2024)
		[arXiv:2409.13048 [hep-th]].
		\href{https://doi.org/10.1103/PhysRevD.109.065023}%
		{\texttt{doi:10.1103/PhysRevD.109.065023}}.

\bibitem{FG_Neumann} C.~D.~Fosco and B.~C.~Guntsche,
		``Quantum dissipative effects for a real scalar field 
		coupled to a dynamical Neumann surface in $d+1$ 
		dimensions,'' Phys.\ Rev.\ D \textbf{110}, 085023 (2024)
		[arXiv:2510.01992 [hep-th]].
		\href{https://doi.org/10.1103/PhysRevD.110.085023}%
		{\texttt{doi:10.1103/PhysRevD.110.085023}}.

\bibitem{GR} I.~S.~Gradshteyn and I.~M.~Ryzhik,
		\textit{Table of Integrals, Series, and Products},
		8th ed., Academic Press (2015).
		\href{https://doi.org/10.1016/C2010-0-64839-5}%
		{\texttt{doi:10.1016/C2010-0-64839-5}}.

\bibitem{FLM_2007} C.~D.~Fosco, F.~C.~Lombardo 
		and F.~D.~Mazzitelli,
		``Quantum dissipative effects in moving mirrors: 
		A functional approach,''
		Phys.\ Rev.\ D \textbf{76}, 085007 (2007)
		[arXiv:0705.2960 [hep-th]].
		\href{https://doi.org/10.1103/PhysRevD.76.085007}%
		{\texttt{doi:10.1103/PhysRevD.76.085007}}.

\bibitem{FLM_2011} C.~D.~Fosco, F.~C.~Lombardo 
		and F.~D.~Mazzitelli,
		``Quantum dissipative effects in moving imperfect mirrors: 
		sidewise and normal motions,''
		Phys.\ Rev.\ D \textbf{84}, 025011 (2011)
		[arXiv:1105.2745 [hep-th]].
		\href{https://doi.org/10.1103/PhysRevD.84.025011}%
		{\texttt{doi:10.1103/PhysRevD.84.025011}}.

\end{thebibliography}
\end{document}